# Spatiotemporal Imaging of Thickness-Induced Band Bending Junctions

*Joeson Wong[1], Artur R. Davoyan[1,2], Bolin Liao[3,4], Andrey Krayev[5], Kiyoung Jo[6], Eli Rotenberg[7], Aaron Bostwick[7], Chris Jozwiak[7], Deep Jariwala[1,6], Ahmed Zewail[3, ‡], Harry A. Atwater[1, *]*

1. Department of Applied Physics and Materials Science, California Institute of Technology, Pasadena, CA, USA
2. Department of Mechanical and Aerospace Engineering, University of California, Los Angeles, CA, USA
3. Department of Chemistry, California Institute of Technology, Pasadena, CA, USA
4. Department of Mechanical Engineering, University of California, Santa Barbara, CA, USA
5. Horiba Scientific, Novato, CA, USA
6. Department of Electrical Engineering, University of Pennsylvania, Philadelphia, PA, USA
7. Advanced Light Source, Lawrence Berkeley National Lab, Berkeley, CA, USA

‡ Deceased

*Corresponding Author: Harry A Atwater (haa@caltech.edu)

**ABSTRACT:**

Van der Waals materials exhibit naturally passivated surfaces and can form versatile heterostructures, enabling observation of carrier transport mechanisms not seen in three-dimensional materials. Here we report observation of a "band bending junction", a new type of semiconductor homojunction whose surface potential landscape depends solely on a difference in thickness between the two semiconductor regions atop a buried heterojunction interface. Using $MoS_2$ on Au to form a buried heterojunction interface, we find that lateral surface potential differences can arise in $MoS_2$ from the local extent of vertical band bending in thin and thick $MoS_2$ regions. Using scanning ultrafast electron microscopy, we examine the spatiotemporal dynamics of photogenerated charge carriers and find that lateral carrier separation is enabled by a band bending junction, which is confirmed with semiconductor transport simulations. Band bending junctions may therefore enable new electronic and optoelectronic devices in Van der Waals materials that rely on thickness variations rather than doping to separate charge carriers.

## INTRODUCTION

Band bending in semiconductors is a fundamental consequence of incomplete screening of external fields and is critical to the operation of nearly every electronic and optoelectronic device. Its existence was theoretically proposed by Mott and Schottky[1,2], who argued that the electrostatic landscape must have electronic bands that "bend" to compensate the difference in Fermi levels at an interface to minimize the overall free energy in the system. Mott also discovered that a characteristic length scale for the band bending in semiconductors is given by

$$L_D = \sqrt{\frac{\epsilon_S \epsilon_0 kT}{q^2 \rho_0}} \tag{1}$$

which is now commonly referred to as the Debye screening length, after Peter Debye who discovered the same parameter in electrolytes[3]. The screening length $L_D$ usually ranges from 10s of nm to a few microns, depending on the doping concentration $\rho_0$ and static dielectric constant $\epsilon_S$. Here, $\epsilon_0$ is the permittivity of free space, $kT$ is the thermal energy, and $q$ is the fundamental unit of charge.

Layered van der Waals materials such as the semiconducting transition metal dichalcogenides (TMDCs) exhibit distinctive band bending physics because of their highly passivated surfaces and the ability to form a wide assortment of heterostructures, which has enabled applications including transistors, solar cells, optical modulators, metasurfaces, and lasers[4–9]. Furthermore, these materials can be easily cleaved to yield layers over a wide range of thicknesses, ranging from a single monolayer to 100s of nm.

Despite considerable research on layered materials in the atomically thin limit in recent years, there has been to our knowledge no direct observation of 'vertical' band bending (i.e., in the direction perpendicular to heterostructure interfaces). This is primarily due to the weak out of plane as opposed to in-plane transport in layered materials and the difficulty to probe buried interfaces. Meanwhile, direct observation of band bending can be used to estimate depletion widths, interface barrier heights, and consequently be used to deduce the electrostatic landscape and performance of the corresponding device.

In this paper, we show evidence for direct measurement of vertical band bending in a $MoS_2$-Au interface. We directly observe correlations between the thickness and surface Fermi levels in samples with identical electronic band structures and preparation methods and find that the $MoS_2$-Au interface results in a strong electron transfer to the $MoS_2$ layer. The direct observation of a surface potential difference between materials with differing thicknesses suggests that a new type of homojunction, arising solely from the differences in thickness and band bending, can be used to separate charge carriers. We directly observe this charge carrier separation by utilizing scanning ultrafast electron microscopy and Kelvin probe microscopy and corroborate these observations with numerical simulations. The electrostatic landscape of materials that are comparable to or thinner than their electrostatic screening length can therefore be carefully tailored by control of their thicknesses, interfaces, and local geometry.

## RESULTS

### Correlation between Electronic Properties and Thickness in Ultrathin Semiconductors

To examine the interplay between interfaces, thicknesses, and band bending, we first consider theoretically a semiconductor on a metallic substrate surrounded by vacuum and solve Poisson's equation

$$\nabla^2 \phi = -\frac{\rho}{\epsilon} \tag{2}$$

to calculate the energy band diagrams for differing thicknesses, as depicted in Figure 1. As expected, we find the characteristic length scale to be the Debye screening length $L_D$, which we estimate to be approximately 40 nm for carrier concentrations corresponding to typical values for MoS2. Furthermore, our calculations suggest that layers that are thinner than their Debye screening length will have a surface potential that is directly related to its thickness. Importantly, this analysis requires use of a material, like MoS2, with an absence of surface states and other contaminants at the top (basal-plane) semiconductor-vacuum interface that is typical in most three-dimensional materials due to the formation of dangling bonds. Van der Waals materials are therefore ideal for probing this thickness-dependent surface potential because of their naturally passivated surfaces.

To examine the effects of varying the semiconductor thickness on surface potential, we directly exfoliated MoS2 on Au (see Methods for details). These exfoliated samples produce a variety of thicknesses that can be determined with atomic force microscopy (AFM) and Kelvin Probe Force Microscopy (KPFM), as shown in Figure 2a and Figure 2b. These images clearly show a direct correlation between two different thicknesses of a MoS2 flake (with thicknesses of about 15 nm and 60 nm), with correspondingly different surface potentials (50 mV and −230 mV), as shown in Figure 2c. Importantly, these thicknesses are outside the realm where there are strong quantum confinement effects and therefore can be considered electronically as 'bulk' materials[10]. Since these two flakes were fabricated under the same conditions, we therefore attribute the difference in observed surface potentials to vertical band bending at the MoS2-Au interface. We further show this correlation between thickness and surface potential for a variety of thicknesses measured on other samples (See Supplementary Figure S1).

The relative surface potential difference between the 15 nm and 60 nm sample is about 280 mV, suggesting strong electron transfer from the gold substrate to the ultrathin MoS2 (Figure 2d). Strong electron transfer at the MoS2-Au interface appears contradictory to the well-known work function of Au (~5.1 eV), which instead would suggest hole doping of the neighboring MoS2 layer. However, recent works has shown that the MoS2-Au interface induces strong electron transfer[11,12], particularly if the interface remains pristine during the exfoliation process, which has also enabled large area monolayer exfoliation of TMDCs and other 2D materials[12–16].

To further examine this correlation between thickness and electronic properties, we perform spatially resolved angle-resolved photoemission spectroscopy (ARPES) at Beamline 7.0.2 at the Advanced Light Source (see methods) on another flake of $MoS_2$ on Au (Figure 2e). We find a direct correlation between the position of the sulfur 2p core levels and the thicknesses of the corresponding $MoS_2$ flakes (Figure 2f, g). The larger sulfur 2p binding energies for the thinner $MoS_2$ is suggestive of electron transfer at the $MoS_2$-Au interface. Finally, we examine electron binding energies and momenta that correspond to the valence band edge, which occurs at the $\Gamma$ point in Brillouin zone in electronically bulk samples. We find that while the thicker sample has a Fermi level of about 1 eV above the valence band edge, the Fermi level at the surface of the thinner sample is about 1.3 eV above its valence band edge (Figure 2h, i). Assuming an electronic bandgap of approximately 1.3 eV[17], we find direct evidence of strong electron transfer and vertical band bending at the $MoS_2$-Au interface, which corroborates the KPFM results shown earlier. Similar results are also observed on another $MoS_2$ sample (See Supplementary Figure S2). Given that the native doping of bulk $MoS_2$ is typically n-type, these results suggest the $MoS_2$-Au interface would yield Ohmic n-type contacts by degenerately doping the neighboring $MoS_2$ region[18].

**Spatiotemporal Imaging of Charge Carrier Separation due to Thickness**

To investigate the effects of the different vertical band bending profiles on the charge carrier dynamics, we perform scanning ultrafast electron microscopy (SUEM) on another sample of $MoS_2$/Au with a thin-thick junction of 10 nm and 100 nm, respectively (Figure 3b). Figure 3a shows a conceptual depiction of the SUEM measurement technique performed over this thin-thick junction. Briefly, SUEM is a pump-probe technique that uses an optical pump ($\sim$514 nm) and an electron pulse ($\sim$2 ps pulse width) that records the secondary electron emission with the presence of the optical pump as a function of pump-probe time delay. Contrast images can be formed at different time delays by subtracting the static SEM image (Figure 3c) from a similar SEM image formed with the optical pump on at a specific time delay $\Delta t$. The contrast in the secondary electron emission as a result of the optical pump beam has been interpreted as direct images of electron (blue) and hole (red) carrier populations under ultrafast excitation, which has been previously used to image carrier separation in pn junctions[19]. Similar methods to directly image ultrafast carrier dynamics has also been utilized in photoemission electron microscopy[20,21].

By examining the contrast images formed at different time delays via SUEM, we find direct evidence for carrier separation at a thin-thick junction (Figure 3d). First, the lack of carrier dynamics at negative time delays suggests appropriate background subtraction. At longer time delays, we initially find the appearance of excess holes (red contrast) on every thickness of $MoS_2$ for $\Delta t < 27$ ps. The oblong contrast profile is due to the beam shape of oblique excitation. At $\Delta t = 33$ ps, there is a simultaneous occurrence of both excess holes and electrons on the thin (10 nm) sample, with the excess electrons located near the thin-thick junction. At slightly longer time delays (40 ps $< \Delta t <$ 60 ps), this population of excess electrons appears to increase and spread before monotonically decreasing with a similar time constant (single exponential fit yields $\tau \sim 75$ ps) to that of the excess hole population on the thick $MoS_2$ ($\Delta t > 60$ ps). Interestingly, the intermediate thickness of $MoS_2$ (light green outline, Figure 3d), yields little to no excess electrons, suggesting the dominant path for carrier transport is between the thin (10 nm) and thick (100 nm) layers of MoS$_2$. Further measurements on monolayer MoS$_2$ and bulk (>>100 nm thick) MoS (See Supplementary Figure S3 and S4) yield little carrier dynamics and no appearance of excess electrons, as observed at this thin-thick junction.

We interpret the carrier dynamics in these contrast images as direct evidence of carrier separation at a thin-thick layer interface, which we refer to hereafter as a "band bending junction". The dynamics can be qualitatively explained as follows (See Supplementary Figure S5 for a schematic): (1) optical excitation results in generation of electron-hole pairs, which rapidly separate vertically due to vertical band bending and thicknesses in the semiconductor that are small compared to carrier diffusion lengths. The strong electron transfer at the MoS$_2$-Au interface results in a band profile schematically depicted in Figure 1, which causes holes (electrons) to move toward the MoS$_2$-vacuum (MoS$_2$-Au) interface for both thicknesses. (2) After the carriers are separated, both carriers can travel on average over one diffusion length within their lifetime. Therefore, if the holes on the thin side are within a diffusion length of the thin-thick junction, they can travel laterally and vertically to separate from the remaining electrons on the thin side. Electrons can also travel within their own diffusion length but will remain near the MoS$_2$-Au interface due to the band profile. (3) Once some holes move across the thin-thick interface, remaining holes follow along due to the gradient in the quasi-Fermi level in the holes. (4) After the electrons and holes have completely separated, they recombine primarily monomolecularly through trap states.

To model the lateral carrier separation dynamics of this band bending junction, we turn to time-dependent drift-diffusion equations (see methods). To simplify the numerical modelling substantially, we make the following assumptions: (1) The MoS$_2$-Au interface results in strong electron transfer to MoS$_2$. (2) The dynamics can be qualitatively modelled by semiconductor drift-diffusion equations. Ab-initio calculations combined with the Boltzmann transport equations may yield more accurate results while considering the unique bandstructures of these systems. (3) the carrier dynamics can be effectively modelled in two dimensions, due to the prominent dynamics at the thin (10 nm) - thick (100 nm) interface. (4) The anisotropic mobilities in the vertical vs. horizontal directions is effectively compensated by scaling the horizontal dimension by the ratio of the square root of the mobilities. (5) While the carrier dynamics in other semiconductor materials at these time scales has been shown to exhibit super diffusion and therefore a time-dependent mobility[22,23], we fix our mobility to a single value which represents effectively a time and spatially averaged quantity. (6) Since we are interested in primarily the lateral carrier dynamics, we assume carrier excitation occurs at 27 ps in the simulation. (7) Secondary electron contrast is primarily due to the net charge density near the surface of the MoS$_2$ layer, with the secondary electron emission decaying exponentially from the surface of the semiconductor with a length scale $\lambda_{SE}$.

With these simulation assumptions in mind, we find excellent agreement between the experimental secondary electron emission contrast and the calculated time-dependent net negative surface charge density (Figure 4a). Snapshots of the charge density at specific times are also shown in Figure 4b, showing lateral carrier separation at the thin-thick interface shortly after excitation. The agreement has been achieved from a variety of simplifications in the theoretical modelling, which suggests that the extracted material parameters used to achieve this matching (specifically, the electron and hole mobility of $MoS_2$) should not be taken to be necessarily physical. However, we find that the qualitative picture of the band bending junction is robust (See Supplementary Figure S6) to a variety of material constants, i.e., lateral carrier separation is achieved independent of the specific choice of material parameters, suggesting that this junction should be observable in other van der Waals materials as well. Our calculations also suggest the current density at the thin-thick interface is dominated by the hole current (See Supplementary Figure S7), which is expected since it acts as a minority carrier in this electron-doped system.

**CONCLUSIONS**

Our results suggest that the interplay between thickness and band bending for materials thinner than their screening length can result in the formation of a new type of homojunction, which we refer to as a band bending junction. These band bending junctions are formed by combining the highly passivated surfaces of van der Waals materials with vertical band bending in materials whose thicknesses are comparable to or below the characteristic electrostatic screening length. Thus, while these results were obtained with the $MoS_2$/Au system, they are likely generalizable to other van der Waals heterostructures and perhaps also 3D bonded semiconductors if one can generate chemically passivated yet electronically active surfaces in the ultrathin thickness (<100 nm) limit. Furthermore, these band bending junctions may find use as new photodetector geometries or could be useful for sensing applications. From the fundamental perspective, they may enable the formation of two-dimensional electron/hole gases or enable a wide variety of surface-sensitive measurement techniques to indirectly examine vertical carrier transport in layered van der Waals materials. More generally, the observation of charge carrier transport at these band bending junctions should be considered while modelling and interpreting ultrathin optoelectronic devices that are geometrically inhomogeneous.

## METHODS

### Sample Preparation

Large area ultrathin flakes of $MoS_2$ were fabricated utilizing the strength of the Au-S bonds, which has been known to yield large area samples[12–14]. To summarize our procedure, we first created atomically smooth Au substrates using template stripping techniques which routinely yields <3 Å RMS[5,24]. Then, bulk crystals of $MoS_2$ (HQgraphene) were cleaved from the native crystal using thermal release tape (Semiconductor Equipment Corp., No. 3195MS) and directly pressed onto the template stripped Au substrate. With the tape and bulk $MoS_2$ directly pressed onto the Au substrate, the entire stack was then heated to ~120 C on a hot plate for ~2 minutes to release the bulk crystal from the tape and to promote adhesion between the $MoS_2$ and Au substrate. The $MoS_2$/Au sample was then sonicated in Acetone for ~5 seconds to release the bulk crystal from the ultrathin layer of $MoS_2$ that would remain adhered to the Au substrate. The sample was finally rinsed with Isopropanol and blow dried with $N_2$. We found it was necessary to template strip the Au substrate immediately before pressing with the $MoS_2$ bulk crystal to yield large area flakes, similar to what has been observed previously[12].

### Kelvin Probe Force Microscopy Measurements

Scanning probe microscopy was performed on OmegaScope-R SPM (AIST-NT, now-Horiba Scientific). HQ NSC-14-Cr/Au probes (Mikromasch) were used for characterization. Kelvin probe imaging was performed in frequency modulation mode which allowed improved spatial resolution of the distribution of the contact potential difference (which reflects the distribution of the surface potential on the sample). The value of the surface potential of the probes was not calibrated, so it was the contrast in the CPD images, not the absolute value of the surface potential, which bore the physical meaning in the CPD images.

### Photoemission Spectroscopy Measurements

Photoemission spectroscopy measurements were performed at Beamline 7.0.2 (MAESTRO) at the Advanced Light Source. Samples were characterized at the microARPES UHV endstation, with synchrotron beam spot sizes of approximately 10 $\mu m$. Incident photon energies were 145 eV and 330 eV for the valence band and core level measurements, respectively. Measurements were performed at ~72 K and multiple frames were averaged together to achieve sufficient signal to noise ratios.

### Scanning Ultrafast Electron Microscopy Measurements

Scanning ultrafast electron microscopy is a newly developed technique that can directly image the dynamics of photoexcited carriers in both space and time with subpicosecond temporal resolution and nanometer spatial resolution. Details of the setup can be found elsewhere[25,26] and are briefly summarized here (also illustrated in Figure 3a). Compared to optical pump−probe spectroscopy, SUEM is a photon−pump−electron−probe technique, with subpicosecond electron pulses generated by illuminating a photocathode (ZrO-coated tungsten tip) with an ultrafast ultraviolet (UV) laser beam (wavelength 257 nm, pulse duration 300 fs, repetition rate 5 MHz, fluence 300 $\mu J/cm^2$). A typical probing electron pulse consists of tens to hundreds of electrons, estimated by measuring the beam current through a Faraday cup, and is accelerated to 30 keV before impacting the sample. The probing electron pulses arrive at the sample after the optical pump pulses (wavelength 515 nm, fluence 80 $\mu$J/cm$^2$) by a given time controlled by a mechanical delay stage ($-700$ ps to 3.6 ns with 1 ps resolution). The probing electron pulses induce the emission of secondary electrons from the sample, which are subsequently collected by an Everhart−Thornley detector. To form an image, the probing electron pulses are scanned across the sample surface and the secondary electrons emitted from each location are counted. Because the yield of secondary electrons depends on the local average electron energy, more/less secondary electrons are emitted from regions of the sample surface where there is a net accumulation of electrons/holes. Typically, a reference SEM image is taken long before the pump optical pulse arrives and is then subtracted from images taken at other delay times to remove the background. In the resulting "contrast images", blue/red contrasts are observed at places with net accumulation of electrons/holes due to higher/lower yield of secondary electrons. In this fashion, the dynamics of electrons and holes after excitation by the optical pump pulse can be monitored in real space and time.

**Electromagnetic and Transport Simulations**

The coupled Drift-Diffusion and Poisson equations were solved using the CHARGE solver in Lumerical DEVICE, which uses finite-element meshing to solve the coupled differential equations iteratively. Due to the anisotropic mobility known for these materials[27,28], solving the multi-dimensional coupled differential equations is computationally costly. Instead, we argue that since we are interested in a qualitative picture of the carrier transport and because the lateral extent of the junction is dominated by the in-plane diffusion length, a proper rescaling of the in-plane dimensions by a factor of $\frac{L_\parallel}{L_\perp} = \sqrt{\frac{\mu_\parallel}{\mu_\perp}} \sim \sqrt{1000}$ allows us to treat the problem with an approximate isotropic, spatially, and temporally averaged mobility ($\mu = 2000$ cm$^2$/(V-s)). Since the experimentally observed in-plane junction width is $\sim$10 $\mu$m, the rescaled in-plane width should be about 300 nm. In our simulations our total lateral span is 500 nm. Band bending was captured assuming an Ohmic contact at the Au-MoS$_2$ interface with a metal work-function of 3.8 eV and semiconductor electron affinity of 4.0 eV. Fermi-Dirac statistics were included due to the high level of modulation doping. Furthermore, we used a bandgap value of 1.3 eV[17], hole effective mass of $0.785m_0$[29], an electron effective mass of $0.686m_0$ (for the electron effective mass, we took the geometric mean of the effective masses in the transverse and longitudinal directions at the Q point[29]), dielectric constant of 7[30], and a Shockley-Read-Hall recombination lifetime of 75 ps. Exciton dynamics were deemed not relevant, since majority of the transport should occur at the lowest energy conduction band and highest energy valence band, with exciton binding energies $\ll kT$[31]. Native doping of MoS$_2$ was assumed to be n-type with a doping level of 10$^{16}$ cm$^{-3}$, similar to that quoted from the supplier. Optical generation values were calculated assuming the system is optically one-dimensional over a specific thickness (either 10 nm or 100 nm), and therefore 1D transfer matrix calculations were applicable for each region. These generation rates were then directly imported into Lumerical DEVICE. Calculated volumetric charge densities were exponentially weighted from the surface with a characteristic length scale of $\lambda_{SE} = 4$ nm to yield surface charge densities. The SUEM signal is expected to be proportional to the net negative charge density $I_{SUEM} \propto -\delta\rho = -q(\delta p - \delta n)$, where positive/negative SUEM signal scales with the net electron/hole population.

**ASSOCIATED CONTENT:**

**AUTHOR INFORMATION**

**Corresponding Author:**

*Harry A. Atwater, E-mail: haa@caltech.edu

**ORCID:**

Joeson Wong: 0000-0002-6304-7602

Artur Davoyan: 0000-0002-4662-1158

Bolin Liao: 0000-0002-0898-0803

Kiyoung Jo: 0000-0003-4587-234X

Eli Rotenberg: 0000-0002-3979-8844

Aaron Bostwick: 0000-0002-9008-2980

Chris Jozwiak: 0000-0002-0980-3753

Deep Jariwala: 0000-0002-3570-8768

Harry Atwater: 0000-0001-9435-0201

**Author Contributions:**

J.W, A.R.D, D.J, and H.A.A developed the main ideas. J.W, A.R.D, and D.J fabricated the samples. B.L performed SUEM measurements, developed in the lab of A.Z. A.K and K.J performed the Kelvin Probe measurements. J.W performed ARPES measurements with support from D.J, E.R, A.B, and C.J. J.W performed both the time-domain and steady-state simulations with assistance from A.R.D. H.A.A supervised over all the data analysis. All authors contributed to the discussion and interpretation of results, as well as the presentation and preparation of the manuscript.

**Notes:**

The authors declare no competing financial interests.

**ACKNOWLEDGEMENTS:**

This work was primarily supported by the 'Photonics at Thermodynamic Limits' Energy Frontier Research Center funded by the U.S. Department of Energy, Office of Science, Office of Basic Energy Sciences under Award Number DE-SC0019140, which supported the sample fabrication, experimental measurements, data analysis, and simulations. J.W acknowledges additional support from the National Science Foundation Graduate Research Fellowship under Grant No. 1144469. Data analysis by A.R.D acknowledges support from UCLA Council on Research Faculty Research Grant. Data analysis by B. L acknowledges support for this work from the U.S. Army Research Office under the award number W911NF-19-1-0060. Additional Kelvin probe force microscopy measurements and data analysis, performed by D.J and K.J, acknowledge support for this work by the U.S. Army Research Office under contract number W911NF-19-1-0109. The work at Penn was carried out at the Singh Center for Nanotechnology which is supported by the National Science Foundation (NSF) National Nanotechnology Coordinated Infrastructure Program grant NNCI-1542153. This research used resources of the Advanced Light Source, a U.S. DOE Office of Science User Facility under contract no. DE-AC02-05CH11231.

**REFERENCES:**

1. Schottky, W. Zur halbleitertheorie der Sperrschicht- und Spitzengleichrichter. *Zeitschrift für Phys. Vol.* **113**, 367–414 (1939).

2. Mott, N. F. Note on the contact between a metal and an insulator or semi-conductor. *Proc. Cambridge Philos. Soc.* **34**, 568–572 (1938).

3. Debye, P. & Hückel, E. Zur Theorie der Elektrolyte. I. Gefrierpunktserniedrigung und verwandte Erscheinungen. *Phys. Zeitschrift* **24**, 185–206 (1923).

4. Jariwala, D., Sangwan, V. K., Lauhon, L. J., Marks, T. J. & Hersam, M. C. Emerging device applications for semiconducting two-dimensional transition metal dichalcogenides. *ACS Nano* **8**, 1102–1120 (2014).

5. Wong, J. *et al.* High photovoltaic quantum efficiency in ultrathin van der Waals heterostructures. *ACS Nano* **11**, (2017).

6. Datta, I. *et al.* Low-loss composite photonic platform based on 2D semiconductor monolayers. *Nat. Photonics* **14**, 256–262 (2020).

7. van de Groep, J. *et al.* Exciton resonance tuning of an atomically thin lens. *Nat. Photonics* **14**, 426–430 (2020).

8. Paik, E. Y. *et al.* Interlayer exciton laser of extended spatial coherence in atomically thin heterostructures. *Nature* **576**, 80–84 (2019).

9. Radisavljevic, B., Radenovic, A., Brivio, J., Giacometti, V. & Kis, A. Single-layer $MoS_2$ transistors. *Nat. Nanotechnol.* **6**, 147–50 (2011).

10. Zhao, W. *et al.* Origin of indirect optical transitions in few-layer $MoS_2$, $WS_2$, and $WSe_2$. *Nano Lett.* **13**, 5627–5634 (2013).

11. Velický, M. *et al.* Strain and charge doping fingerprints of the strong interaction between monolayer $MoS_2$ and gold. *J. Phys. Chem. Lett.* **11**, 6112–6118 (2020).

12. Velický, M. *et al.* Mechanism of gold-assisted exfoliation of centimeter-sized transition metal dichalcogenide monolayers. *ACS Nano* **12**, 10463–10472 (2018).

13. Magda, G. Z. *et al.* Exfoliation of large-area transition metal chalcogenide single layers. *Sci. Rep.* **5**, 14714 (2015).

14. Desai, S. B. *et al.* Gold-mediated exfoliation of ultralarge optoelectronically-perfect monolayers. *Adv. Mater.* **28**, 4053–4058 (2016).

15. Liu, F. *et al.* Disassembling 2D van der Waals crystals into macroscopic monolayers and reassembling into artificial lattices. *Science* **367**, 903–906 (2020).

16. Huang, Y. *et al.* Universal mechanical exfoliation of large-area 2D crystals. *Nat. Commun.* **11**, (2020).

17. Mak, K. F., Lee, C., Hone, J., Shan, J. & Heinz, T. F. Atomically thin $MoS_2$: A new direct-gap semiconductor. *Phys. Rev. Lett.* **105**, 2–5 (2010).

18. Jariwala, D. *et al.* Band-like transport in high mobility unencapsulated single-layer MoS$_2$ transistors. *Appl. Phys. Lett.* **102**, 173107 (2013).

19. Najafi, E., Scarborough, T. D., Tang, J. & Zewail, A. Four-dimensional imaging of carrier interface dynamics in p-n junctions. *Science* **347**, 164–167 (2015).

20. Man, M. K. L. *et al.* Imaging the motion of electrons across semiconductor heterojunctions. *Nat. Nanotechnol.* **12**, 36–40 (2017).

21. Wong, E. L. *et al.* Pulling apart photoexcited electrons by photoinducing an in-plane surface electric field. *Sci. Adv.* **4**, (2018).

22. Najafi, E., Ivanov, V., Zewail, A. & Bernardi, M. Super-diffusion of excited carriers in semiconductors. *Nat. Commun.* **8**, 1–7 (2017).

23. Liao, B. *et al.* Spatial-temporal imaging of anisotropic photocarrier dynamics in black phosphorus. *Nano Lett.* **17**, 3675–3680 (2017).

24. Vogel, N., Zielieniecki, J. & Köper, I. As flat as it gets: ultrasmooth surfaces from template-stripping procedures. *Nanoscale* **4**, 3820 (2012).

25. Mohammed, O. F., Yang, D. S., Pal, S. K. & Zewail, A. H. 4D scanning ultrafast electron microscopy: visualization of materials surface dynamics. *J. Am. Chem. Soc.* **133**, 7708–7711 (2011).

26. Yang, D. S., Mohammed, O. F. & Zewail, A. H. Scanning ultrafast electron microscopy. *Proc. Natl. Acad. Sci. U. S. A.* **107**, 14993–14998 (2010).

27. Massicotte, M. *et al.* Picosecond photoresponse in van der Waals heterostructures. *Nat. Nanotechnol.* **11**, 42–46 (2016).

28. Li, D. *et al.* Electric-field-induced strong enhancement of electroluminescence in multilayer molybdenum disulfide. *Nat. Commun.* **6**, (2015).

29. Zahid, F., Liu, L., Zhu, Y., Wang, J. & Guo, H. A generic tight-binding model for monolayer, bilayer and bulk MoS$_2$. *AIP Adv.* **3**, (2013).

30. Laturia, A., Van de Put, M. L. & Vandenberghe, W. G. Dielectric properties of hexagonal boron nitride and transition metal dichalcogenides: from monolayer to bulk. *npj 2D Mater. Appl.* **2**, (2018).

31. Cheiwchanchamnangij, T. & Lambrecht, W. R. L. Quasiparticle band structure calculation of monolayer, bilayer, and bulk MoS$_2$. *Phys. Rev. B - Condens. Matter Mater. Phys.* **85**, 1–4 (2012).

**Figures:**

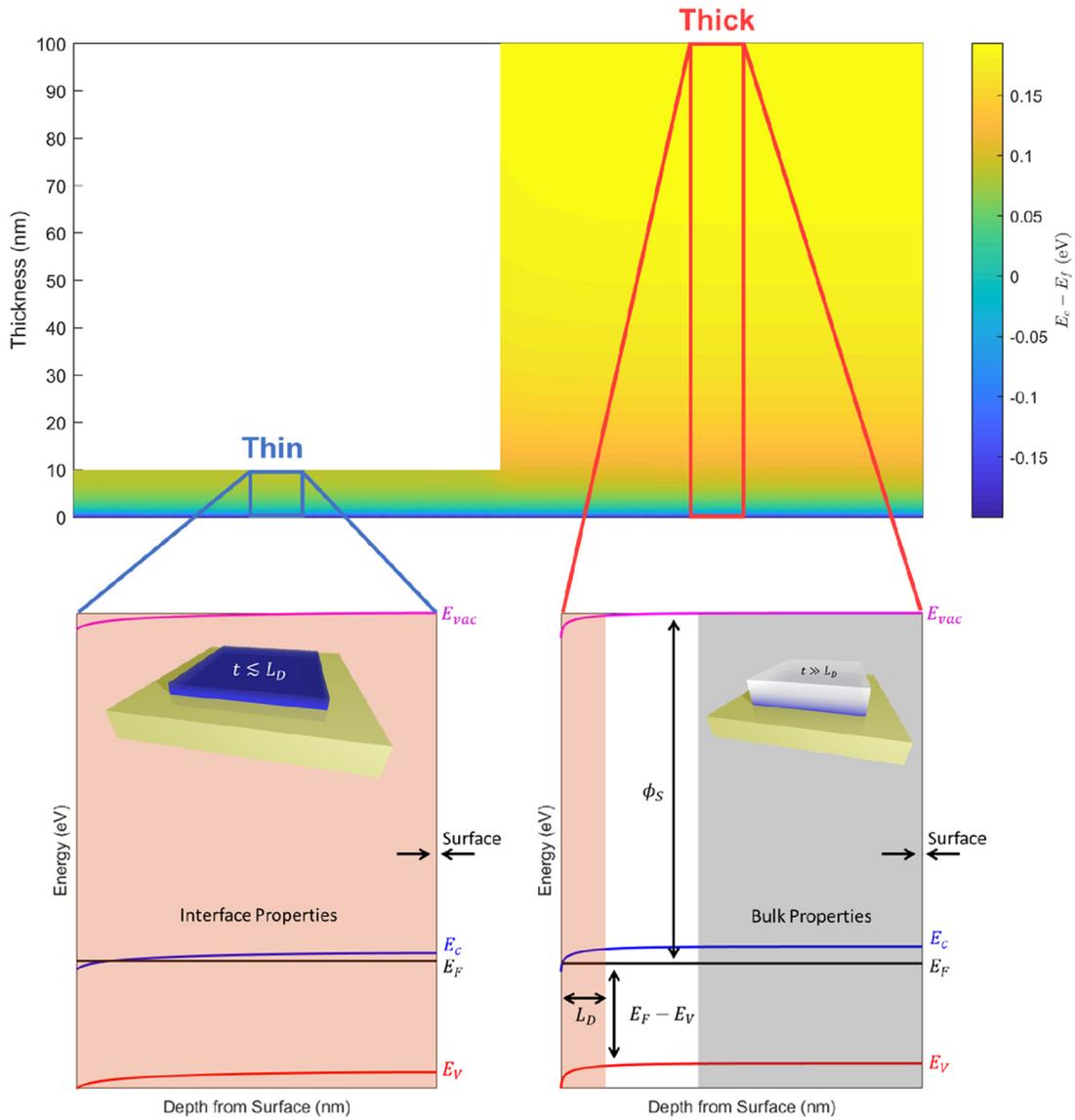

**Figure 1.** <u>Thickness-dependent surface potentials due to vertical band bending:</u> Calculated $E_c - E_f$ band diagram for a 10 nm and 100 nm thick flake of MoS$_2$ on Au assuming strong electron transfer at the MoS$_2$/Au interface. Schematic band diagrams of a material dominated by its interface properties (bottom left) and bulk properties (bottom right), which depends on the thickness of the material relative to its electrostatic screening length ($L_D$). Insets correspond to a schematic of a semiconductor (e.g., MoS$_2$) on its substrate (e.g., Au), with the blue representing excess electron concentration relative to its bulk value. The surface of the material refers to the semiconductor-vacuum interface.

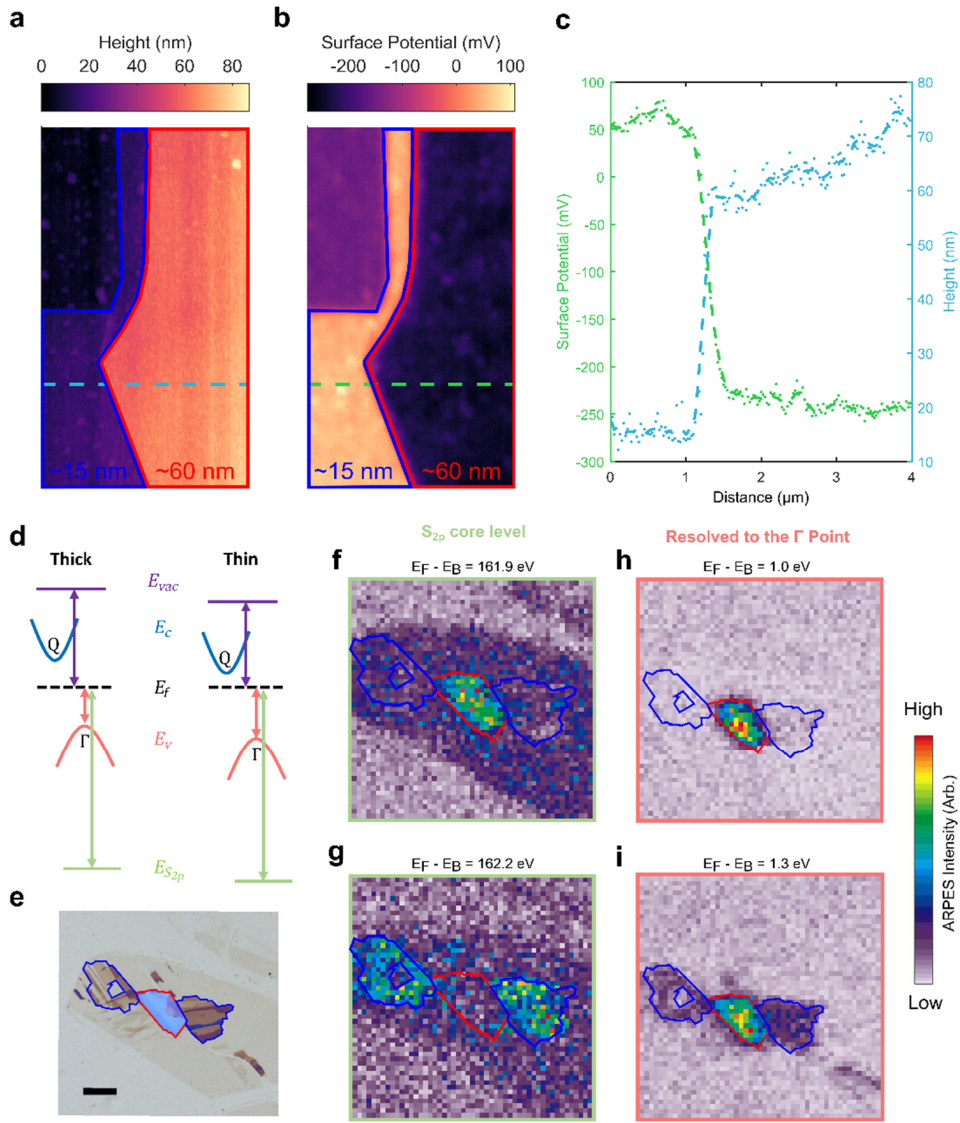

**Figure 2.** <u>Experimental observation of thickness-dependent surface doping in MoS₂/Au:</u> **(a)** Topographic image of MoS₂ exfoliated onto a gold substrate with corresponding surface potential **(b)** mapped over the same area. The blue and red outlines correspond to a MoS₂ thickness of approximately 15 nm and 60 nm, respectively. The upper left region corresponds to monolayer MoS₂/Au. **(c)** Linecut of the topography and surface potential. The dashed lines are guides for the eye. **(d)** Proposed energy diagram at the surface for the thin and thick MoS₂. **(e)** Optical micrograph image of MoS₂ exfoliated onto a gold substrate. The thick (red outline) and thin (blue outline) regions are ~30 nm and ~5 nm thick, respectively (scale bar = 50 $\mu$m). Intensity map of photoemitted electrons at the sulfur 2p core level for binding energies of 161.9 eV **(f)** and 162.2 eV **(g)**. Intensity map of photoemitted electrons from the valence band of MoS₂ resolved to its $\Gamma$ point for binding energies of 1.0 eV **(h)** and 1.3 eV **(i)**. The thin (blue) and thick (red) flake outlines are superimposed.

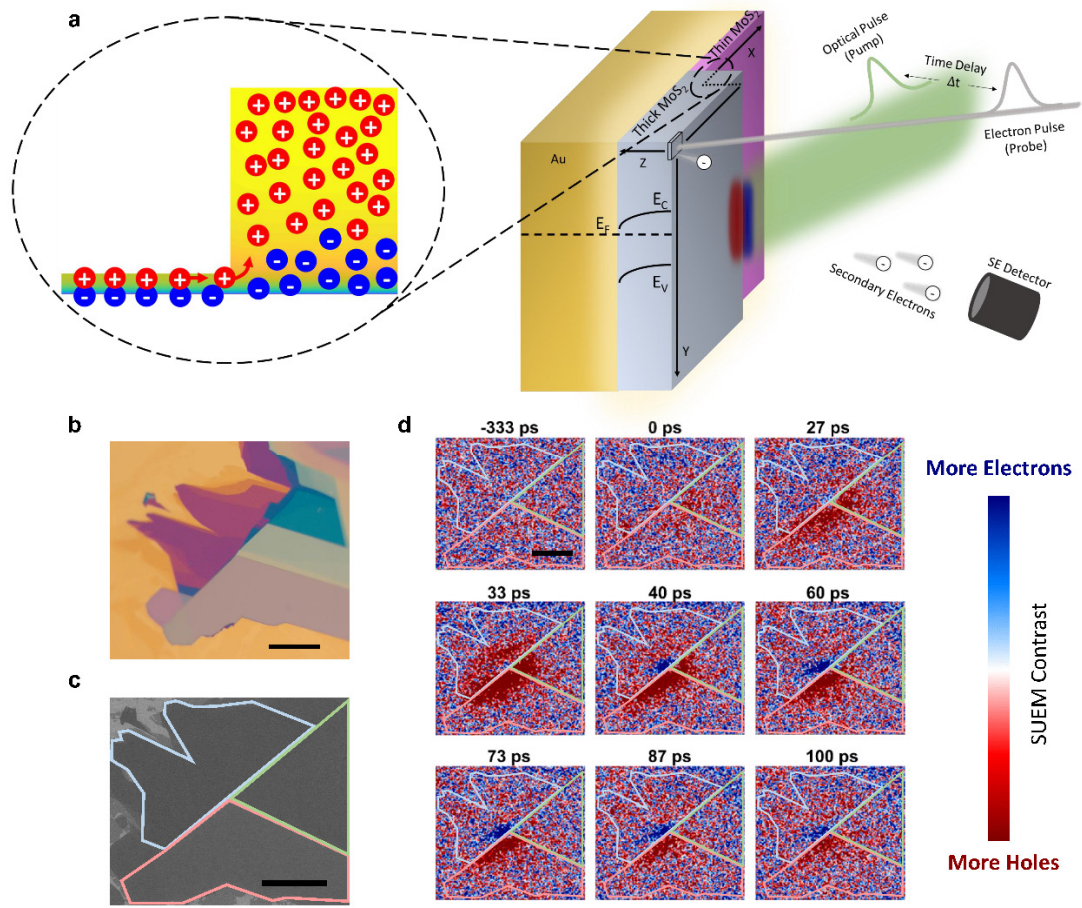

**Figure 3.** <u>Scanning Ultrafast Electron Microscopy Imaging of a Band Bending Junction:</u> **(a)** Conceptual depiction of the band bending junction and its measurement via scanning ultrafast electron microscopy (SUEM). In a sample with different thicknesses of $MoS_2$, band bending induced by a gold substrate enables lateral carrier separation between electrons and holes. In SUEM, an optical pulse generates electron-hole pairs that subsequently evolve in space and time. An electron pulse is raster scanned across the surface of the sample for a given time delay $\Delta t$ after the optical pulse. An image of the detected secondary electron (SE) intensity is formed. Contrast images are formed that correspond to the difference between the SE image at $\Delta t$ relative to the SE image without an optical pulse. Contrast images are interpreted as images of the net charge density, i.e., increased (decreased) SE intensity corresponds to an increase in the local surface electron (hole) density. **(b)** Optical image of $MoS_2$ exfoliated onto a gold substrate. **(c)** Static scanning electron micrograph over the flake in (b), with highlighted regions of thick (pink border, ~100 nm), intermediate thickness (light green border, ~30 nm) and thin (light blue border, ~10 nm) $MoS_2$ on one sample. **(d)** Contrast images over the same area as (c) for different time delays, with corresponding $MoS_2$ flake outline. Blue and red contrasts are interpreted as excess electrons and holes due to photoexcitation, respectively. All scale bars are 50 $\mu$m.

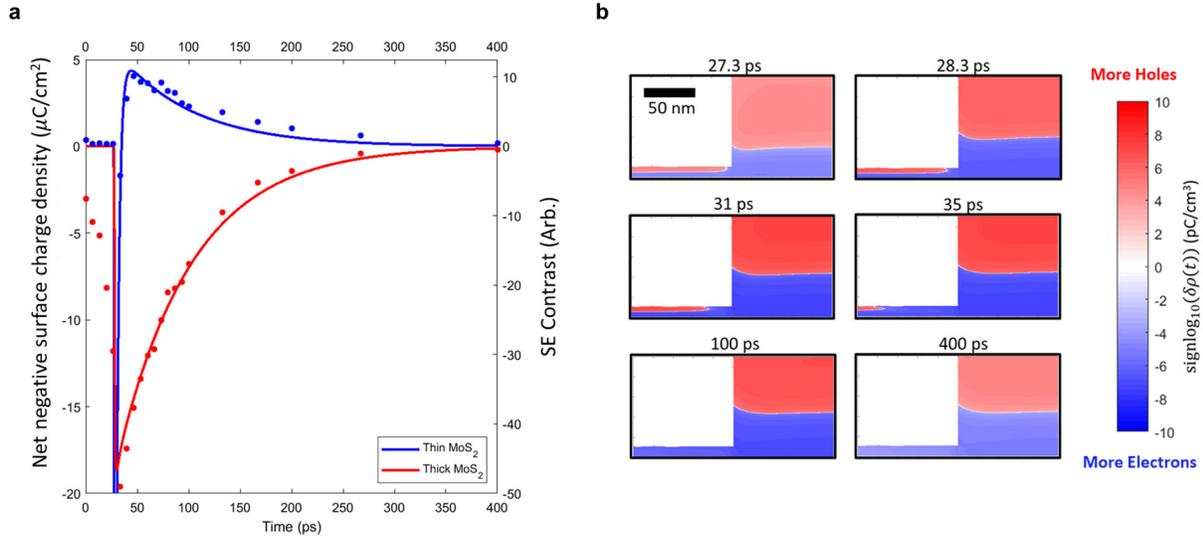

**Figure 4.** <u>Simulation of Carrier Dynamics at a Bend Bending Junction:</u> **(a)** Experimentally measured secondary electron contrast intensity (dots, right axis) on the thin (blue) and thick (red) MoS$_2$ as a function of different pump-probe delay times, along with the simulated net negative surface charge density as a function of time (solid lines, left axis). We assume the thin and thick MoS$_2$ is 10 and 100 nm thick, respectively, for the simulation. **(b)** Simulated cross-sectional maps of the net charge density at different time steps, plotted with a signed log function to examine the orders of magnitude change in carrier density more easily (scale bar = 50 nm). Red corresponds to net positive charge (i.e., holes), while blue corresponds to net negative charge (i.e., electrons).